# Radio Search for H$_2$CCC toward HD 183143 as a Candidate for a Diffuse Interstellar Band Carrier

Short Title: Radio Search for H$_2$CCC

Mitsunori Araki,[1] Shuro Takano,[2] Hiromichi Yamabe,[1] Koichi Tsukiyama,[1] and Nobuhiko Kuze[3]

[1] Department of Chemistry, Faculty of Science Division I, Tokyo University of Science, 1-3 Kagurazaka, Shinjuku-ku 162-8601, Tokyo, Japan; araki@rs.kagu.tus.ac.jp

[2] Nobeyama Radio Observatory and Department of Astronomical Science, The Graduate University for Advanced Studies (Sokendai), 462-2 Nobeyama, Minamimaki, Minamisaku, Nagano 384-1305, Japan

[3] Department of Materials and Life Sciences, Faculty of Science and Technology, Sophia University, 7-1 Kioi-cho, Chiyoda-ku 102-8554, Tokyo, Japan

**ABSTRACT**

To clarify the authenticity of a recently proposed identification of H$_2$CCC (*linear*-C$_3$H$_2$) as a diffuse interstellar band carrier, we searched for the rotational transition of H$_2$CCC at a frequency of 103 GHz toward HD 183143 using a 45-m telescope at the Nobeyama Radio Observatory. Although rms noise levels of 32 mK in the antenna temperature were achieved, detection of H$_2$CCC was unsuccessful, producing a 3 σ upper limit corresponding to a column density of 2.0 × 10$^{13}$ cm$^{-2}$. The upper limit indicates that the contribution of H$_2$CCC to the diffuse interstellar band at 5450 Å is less than 1/25; thus, it is unlikely that the laboratory bands of the $B^1B_1$–$X^1A_1$ transition of H$_2$CCC and the diffuse interstellar bands at 5450 Å (and also 4881 Å) toward HD 183143 are related.

*Subject Keywords*: Astrochemistry — ISM: clouds — ISM: molecules — Radio lines: ISM

## 1. INTRODUCTION

Identification of diffuse interstellar bands (DIBs) is one of the most interesting questions in astrochemistry and spectroscopy and has been recognized for a century. The first DIBs appeared as absorption lines with diffuse linewidths in optical spectra of "red" stars having a large reddening $E_{B-V}$ in the background of diffuse clouds in 1922 (Heger 1922). Subsequently, more than 400 DIBs in the near-infrared and optical regions were observed. In the current widely accepted interpretation, DIBs are electronic transitions of interstellar molecules: carbon chains (Douglas 1977) and/or poly-aromatic hydrocarbons (Leger & D'Hendecourt 1985, Crawford et al. 1985, van der Zwet & Allamandola 1985). However, despite a long period of investigations attempting to match laboratory spectra of molecules and astronomically observed DIB spectra, no molecules could be identified so far.

Last year, Maier and coworkers compared laboratory and astronomically observed spectra and suggested that two DIBs at 4881 and 5450 Å can be assigned to the $B^1B_1$–$X^1A_1$ electronic transition of $H_2CCC$ (*linear*-$C_3H_2$, Maier et al. 2011, Stanton et al. 2012). They found agreements not only in the wavelengths of DIBs but also in their bandwidths in the diffuse clouds toward HD 183143 and other stars. The column density was evaluated as $5 \times 10^{14}$ cm$^{-2}$ toward HD 183143, and the excitation temperature was estimated to be 10–60 K using the line profiles of the transitions. If this identification is correct, it is the first answer to the long-standing puzzle of DIBs.

However, we have a question about the huge column density of $H_2CCC$ in their study. In the dark cloud TMC-1, which is rich in carbon-chain molecules, the reported column density ($2.1 \times 10^{12}$ cm$^{-2}$) of this molecule is not very high (Fossé et al. 2001). Although the column density of $H_2CCC$ evaluated by Maier et al. (2011) was obtained in diffuse clouds, it is about two orders of magnitude higher than that of the dark cloud. Oka & McCall (2011) also pointed out that the abundance of $H_2CCC$ in the study of Maier et al. (2011) was high, in contrast to the low abundance of $C_2$ and $C_3$, toward HD 183143. On the other hand, a few molecules in an interstellar cloud can have huge column densities compared with other molecules. For example, the column density of $C_4H$ in the low-mass star-forming region L1527 is $1.0 \times 10^{14}$ cm$^{-2}$ (Araki et al. 2012), although those of other carbon-chain molecules are $10^{11}$–$10^{13}$ cm$^{-2}$ (Sakai et al. 2008).

Krełowski et al. (2011) recently investigated 49 stars for the two DIBs at 5450 and 4882 Å and reported that the strength (central depth: Ac) ratios Ac(5450)/Ac(4882) of the two DIBs vary considerably, although the ratio should be 1.5 (Fig. 1 of Maier et al. 2011) according to the laboratory spectra, where DIB at 4882 Å studied by Krełowski et al. corresponds to that at 4881

Å studied by Maier et al. Krełowski et al. disagreed that the two DIBs can be identified as $H_2CCC$. However, we found that the many clouds with weak DIB intensities in Figure 7 of Krełowski et al. have ratios of approximately 1.5. We therefore hypothesized that the strong DIBs at 5450 and 4881 Å might be blended with transitions of other molecules and the weak DIBs are due to relatively pure transitions of $H_2CCC$.

Between DIBs at 6159 and 6251 Å, the strength ratio, which might be attributed to the $A^1A_2$–$X^1A_1$ transition of $H_2CCC$ (Maier et al. 2011, Achkasova et al. 2006, Birza et al. 2005), is variable in 49 clouds (Krełowski et al. 2011). Another broad feature of the $B^1B_1$–$X^1A_1$ transition of $H_2CCC$ could appear in the 5165–5185 Å region (Maier et al. 2011). However, this feature was not detected in the spectrum of HD 166734 despite the detection of strong DIBs at 4882 and 5450 Å (Krełowski et al. 2011). If the strong DIBs at 5450 and 4881 Å are actually blended as suggested above, the column density of $H_2CCC$ will be smaller than the value reported by Maier et al. Thus, no detection of the transitions at 6159, 6251, and 5165–5185 Å as DIBs of $H_2CCC$ can be consistent.

Discussions of the column density and intensity ratio based on optical observations cannot give clear evidence on the identification of $H_2CCC$ as a DIB carrier. Radio astronomical observations can provide better evidence that the $H_2CCC$ molecule exists in the diffuse cloud because of the higher resolution and lack of intrinsic broadening mechanisms compared to optical spectroscopy. In this paper, we report the search for $H_2CCC$ as a DIB carrier toward HD 183143.

## 2. OBSERVATIONS

The rotational transition $J_{Ka,Kc} = 5_{1,5}$–$4_{1,4}$ of $H_2CCC$ at 102.99238 GHz (Vrtilek et al. 1990) was observed with the Nobeyama Radio Observatory's 45-m telescope[1] on January 15, 2012. The energy of the upper rotational level $5_{1,5}$ in this transition is 13.8 K (9.6 cm$^{-1}$), which is evaluated on the basis of the lowest rotational level $1_{1,1}$ in the ortho species. The population of the $5_{1,5}$ level at excitation temperatures of 10 and 60 K accounts for 4.3 and 2.1%, respectively, of both the ortho and para species. We selected the diffuse clouds in front of the star HD 183143 to search for $H_2CCC$ because DIBs at both 4881 and 5450 Å were detected toward the star, and a high column density of $5 \times 10^{14}$ cm$^{-2}$ was derived using a central depth of 3.0% for DIB at 5450 Å (Maier et al. 2011). We observed the HD 183143 position [α(J2000), δ(J2000)] = (19$^h$ $27^m\ 26^s.56$, 18°17′45.20″). The SIS mixer receiver S100 was used for the present observations; the typical single sideband system temperature was approximately 260 K. The receiver was used in the frequency switching mode with the frequency throws at 2 MHz because the position of an off point cannot be defined owing to the spatially diffuse structure of diffuse clouds. The main beam efficiency was 38% in the 103 GHz region. The beam size (FWHM) of the telescope is 15.3″ at 103 GHz. The telescope pointing was verified by observing the nearby SiO maser source R-Aql every 60 min using the HEMT receiver H40. The typical pointing accuracy was a few arcseconds. A set of high-resolution acousto-optical radio spectrometers having individual bandwidths of 40 MHz and a frequency resolution of 37 kHz was used for the backend. The corresponding velocity resolution is 0.11 km s$^{-1}$ at 103 GHz. The intensity scale was calibrated by the chopper wheel method.

---

[1] The 45-m telescope is operated by the Nobeyama Radio Observatory, a branch of the National Astronomical Observatory of Japan.

## 3. RESULTS AND DISCUSSION

We searched for the rotational transition $J_{Ka,Kc} = 5_{1,5}$–$4_{1,4}$ of H$_2$CCC toward HD 183143. The obtained spectrum is shown in Figure 1. Although an rms noise level of 32 mK in the antenna temperature was achieved, no lines were detected.

Two velocity components of diffuse clouds toward HD 183143 have been reported for H$_3^+$, CH, CH$^+$, and CN (McCall et al. 2002). By using infrared and optical observations, velocities of 7.7 and 23.6 km s$^{-1}$ and linewidths (FWHM) of 7.7 and 4.9 km s$^{-1}$, respectively, for CH have been measured (McCall et al. 2002). The CH integrated intensity of the stronger velocity component at 23.6 km s$^{-1}$ accounts for 60% of the total intensity of the two velocity components. Because DIBs cannot be split into velocity components owing to the broadband structure in the optical region, the reported column density of $5 \times 10^{14}$ cm$^{-2}$ for H$_2$CCC based on optical observations (Maier et al. 2011) would be the sum of the column densities for the two velocity components. To estimate the sum of the column densities from antenna temperatures obtained by radio observations, we assumed the following conditions: (a) H$_2$CCC has two velocity components just as CH, CH$^+$, and CN; (b) velocities and linewidths of H$_2$CCC in the components are equal to those of CH; and (c) integrated intensities of the components have a ratio of 40:60. On the basis of these assumptions, a line profile of the rotational transition of H$_2$CCC toward HD 183143 was simulated for a tenfold intensity of the rms noise level, as shown in Figure 1.

Maier et al. (2011) suggested an excitation temperature between 10 and 60 K for H$_2$CCC toward HD 183143. A value of ortho/para = 3.0 and a permanent dipole moment of 4.162 D (Wu et al. 2010) were used. Assuming the linewidth of CH, Maier's column density and excitation temperature of 10–60 K, we would expect an antenna temperature of approximately 2 K, which is much higher than the achieved rms noise levels in the present observations. In this study, the observed spectrum (3 σ rms noise) yielded upper limits of $1.1 \times 10^{13}$ and $2.0 \times 10^{13}$ cm$^{-2}$ at 10 and 60 K, respectively, for the summed column densities of the two velocity components toward HD 183143. Local thermal equilibrium was assumed in these calculations. The obtained upper limit of the column density of $H_2CCC$ at 60 K toward HD 183143 is half of those of CH and $CH^+$ observed optically (McCall et al. 2002, Thorburn et al. 2003). The upper limit suggests that the contribution of the transition of $H_2CCC$ to DIB at 5450 Å is less than 1/25. Thus, $H_2CCC$ is unlikely to be a candidate for the DIB carrier toward HD 183143.

The upper limit corresponds to a central depth of 0.12% in DIB at 5450 Å toward HD 183143. The central depth is roughly equal to the detection limits (noise level) of the observations of DIBs at 5450 Å toward 49 stars by Krełowski et al. The present hypothesis noted above, i.e., the weak DIBs observed by Krełowski et al. are due to relatively pure transitions of $H_2CCC$, cannot be supported.

At present, we conclude that the huge column density of $H_2CCC$ toward HD 183143 reported by Maier et al. (2011) is unjustified and that the major carrier of DIBs at 5450 and 4881 Å is not $H_2CCC$.

**Acknowledgment**

M.A. thanks the Tokyo Ohka Foundation for the Promotion of Science and Technology for the financial support.

**REFERENCES**

Achkasova, E., Araki, M., Denisov, A., & Maier, J. P. 2006, *Journal of Molecular Spectroscopy*, **237**, 70

Araki, M., Takano, S., Yamabe, H., Koshikawa, N., Tsukiyama, K., Nakane, A., Okabayashi, T., Kunimatsu, A., & Kuze, N. 2012, *ApJ*, **744**, 163

Birza, P., Chirokolava, A., Araki, M., Kolek, P., & Maier, J. P. 2005, *Journal of Molecular Spectroscopy*, **229**, 276

Crawford, M. K., Tielens, A. G. G. M., & Allamandola, L. J. 1985, *ApJ*, **293**, L45

Douglas, A. E. 1977, *Nature*, **269**, 130

Fossé, D., Cernicharo, J., Gerin, M., & Cox, P. 2001, *ApJ*, **552**, 168

Heger, M. L., 1922, *Lick Observatory bulletin*, no. 337, 141

Krełowski, J., Galazutdinov, G., & Kołos, R. 2011, *ApJ*, **735**, 124

Leger, A. & D'Hendecourt, L. 1985, *A&A*, **146**, 81

Maier, J. P., Walker, G. A. H., Bohlender, D. A., Mazzotti, F. J., Raghunandan, R., Fulara, J., Garkusha, I., & Nagy, A. 2011, *ApJ*, **726**, 41

McCall, B. J., Hinkle, K. H., Geballe, T. R., Moriarty-Schieven, G. H., Evans, N. J., II, Kawaguchi, K., Takano, S., Smith, V. V., & Oka, T. 2002, *ApJ*, **567**, 391

Oka, T. & McCall, B. 2011, *Science*, **331**, 293

Sakai, N., Sakai, T., Hirota, T., & Yamamoto, S. 2008, *ApJ*, **672**, 371

Stanton, J. F., Garand, E., Kim, J., Yacovitch, T. I., Hock, C., Case, A. S., Miller, E. M., Lu, Y.-J., Vogelhuber, K. M., Wren, S. W., Ichino, T., Maier, J. P., McMahon, R. J., Osborn, D. L., Neumark, D. M., & Lineberger, W. C., 2012, *J. Chem. Phys.*, **136**, 134312

Thorburn, J. A., Hobbs, L. M., McCall, B. J., Oka, T., Welty, D. E., Friedman, S. D., Snow, T. P., Sonnentrucker, P., & York, D. G. 2003, *ApJ*, **584**, 339

van der Zwet, G. P. & Allamandola, L. J. 1985, *A&A*, **146**, 76

Vrtilek, J. M., Gottlieb, A., Gottlieb, E. W., Killian, T. C., & Thaddeus, P. 1990, *ApJ*, **364**, L53

Wu, Q., Cheng, Q., Yamaguchi, Y., Li, Q., & Schaefer, H. F. 2010, *J. Chem. Phys.*, **132**, 044308

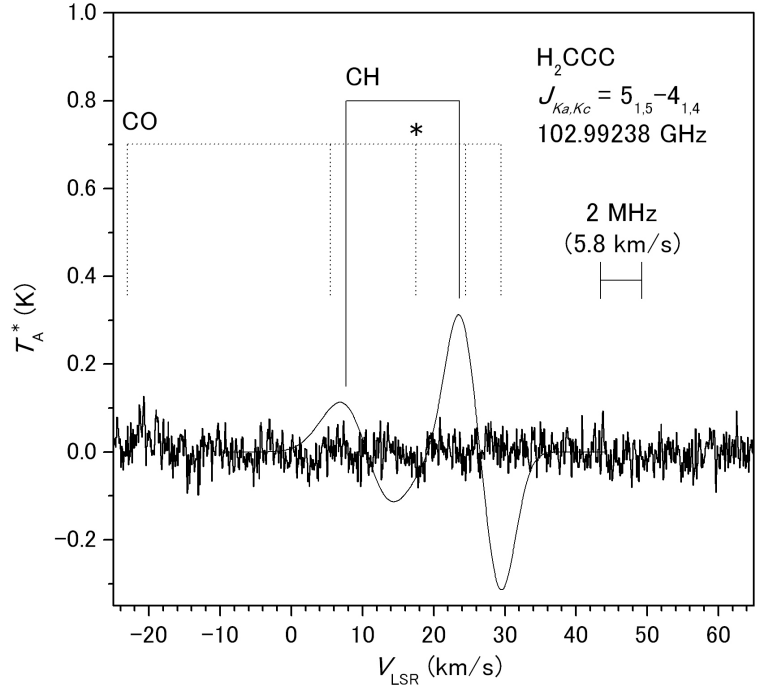

Figure 1. Spectrum at the line position of $H_2CCC$ searched toward HD 183143

Solid vertical lines indicate positions of the two velocity components of CH; the velocity structures of $CH^+$, $H_3^+$, and CN are comparable with that of CH (McCall et al. 2002). Dotted vertical lines indicate positions of the five velocity components of CO by McCall et al.; asterisk marks the strongest component. Solid curve shows a line profile of $H_2CCC$ simulated using the assumptions in the text. A tenfold intensity of the rms noise level in the stronger velocity component at 23.6 km s$^{-1}$ was assumed. In case of the excitation temperature of 60 K, the tenfold intensity corresponds to a column density of $7 \times 10^{13}$ cm$^{-2}$, which is 3.5 times higher than the present upper limit and one order less than Maier's column density. Two downward convex structures in this curve are effects of frequency switching. The intensity of the weaker velocity component at 7.7 km s$^{-1}$ is partially canceled by the downward convex structure.